\documentclass[conference]{IEEEtran}
\IEEEoverridecommandlockouts
\usepackage{cite}
\usepackage{amsmath,amssymb,amsfonts}
\usepackage{algorithmic}
\usepackage{graphicx}
\usepackage{xcolor}

\usepackage{microtype}                 % use micro-typography (slightly more compact, better to read)
\PassOptionsToPackage{warn}{textcomp}  % to address font issues with \textrightarrow
\usepackage{textcomp}                  % use better special symbols
\usepackage{mathptmx}                  % use matching math font
\usepackage{times}                     % we use Times as the main font
\usepackage{cite}                      % needed to automatically sort the references
\usepackage{tabu}                      % only used for the table example
\usepackage{booktabs}                  % only used for the table example

\DeclareMathAlphabet{\mathcal}{OMS}{cmsy}{m}{n}

\usepackage[font={scriptsize,sf}]{caption}

\usepackage{multirow}
\usepackage{paralist}
\usepackage{hyperref}
\usepackage{subcaption}
\usepackage{cleveref}
\usepackage{comment}

\def\BibTeX{{\rm B\kern-.05em{\sc i\kern-.025em b}\kern-.08em
    T\kern-.1667em\lower.7ex\hbox{E}\kern-.125emX}}
\begin{document}

\title{A Visual Analytics Design for Connecting Healthcare Team Communication to Patient Outcomes\\
\vspace{-0.15in}
% \thanks{Identify applicable funding agency here. If none, delete this.}
}

% \author{Submission ID: \#2}
%\\Anonymous Author(s)}
\author{
\IEEEauthorblockN{1\textsuperscript{st} Hsiao-Ying Lu}
\IEEEauthorblockA{\textit{Department of Computer Science} \\
\textit{University of California, Davis}\\
Davis, USA \\
hyllu@ucdavis.edu
\vspace{-0.4in}}
\and
\IEEEauthorblockN{2\textsuperscript{nd} Yiran Li}
\IEEEauthorblockA{\textit{Department of Computer Science} \\
\textit{University of California, Davis}\\
Davis, USA \\
ranli@ucdavis.edu
\vspace{-0.4in}}
\and
\IEEEauthorblockN{2\textsuperscript{rd} Kwan-Liu Ma}
\IEEEauthorblockA{\textit{Department of Computer Science} \\
\textit{University of California, Davis}\\
Davis, USA \\
klma@ucdavis.edu
\vspace{-0.4in}}
% \and
% \IEEEauthorblockN{2\textsuperscript{nd} Given Name Surname}
% \IEEEauthorblockA{\textit{dept. name of organization (of Aff.)} \\
% \textit{name of organization (of Aff.)}\\
% City, Country \\
% email address or ORCID}
}

\maketitle

% \begin{abstract}
\begin{abstract}
Communication among healthcare professionals (HCPs) is crucial for the quality of patient treatment. 
Surrounding each patient's treatment, communication among HCPs can be examined as temporal networks, constructed from Electronic Health Record (EHR) access logs.
This paper introduces EHRFlow, a visual analytics system designed to study the effectiveness and efficiency of temporal communication networks mediated by the EHR system. We present a method that associates network measures with patient survival outcomes and devises effectiveness metrics based on these associations.
To analyze communication efficiency, we extract the latencies and frequencies of EHR accesses. 
EHRFlow is designed to assist in inspecting and understanding the composed communication effectiveness metrics and to enable the exploration of communication efficiency by encoding latencies and frequencies in an information flow diagram. We demonstrate and evaluate EHRFlow through multiple case studies and an expert review.
\end{abstract}

% \end{abstract}


\newcommand{\CaptionStar}{
(a) The star pattern in undirected and unipartite networks. (b) The two-star pattern in directed and bipartite networks. The blue nodes represent one type of nodes and the red nodes represent the other type of nodes. Nodes $i$ and $j$ constitute the two stars in the respective types of nodes.
}

\newcommand{\CaptionTimePaths}{
Illustrations of time-respecting paths: (a) presents an example of a directed temporal graph. The colored nodes are the analysis targets. (b) follows the red highlighted edges, retrieving nodes that can be influenced by node $D$. (c) follows the green highlighted edges, retrieving nodes that can influence node $B$.
}

\newcommand{\CaptionUI}{
(a) and (b) represent two patients with different survival outcomes. (c) is the \textit{metric swarmplot} displaying the communication effectiveness score of every patient in patient group A. (a1) and (b1) are the \textit{stacked boxplots} of the aggregated distance measure for the two patients, with each patient's distance measure depicted using a vertical line atop the aggregated distance stacked boxplot distributions of all patients with different outcomes. (a2) and (b2) illustrate the \textit{measure evolution} of the aggregated distance for the networks of the two patients, brushed with the chosen observation time window. (a3) is the \textit{global information flow diagram}, and (a4) is the \textit{local information flow diagram}, presenting the disseminated set reachable subnetwork of the ego HCP (i.e., the green cross) who treated the survived patient highlighted in (a).
}

\newcommand{\CaptionNIHDataAMetric}{
The rank and metric weights assigned to six representative network measures, providing a metric composition overview for patient group A.
}
\newcommand{\CaptionNIHDataBMetric}{
The rank and metric weights assigned to six representative network measures, providing a metric composition overview for patient group B.
}

\newcommand{\CaptionNIHCaseOne}{
A comparative study of disconnected communications between two representative patient networks selected in \autoref{fig:ui}(a) and \autoref{fig:ui}(b). (a1) and (a2) accentuate the reviewed and disseminated reachable subnetworks of the survived patient, while (b1) and (b2) highlight the reviewed and disseminated reachable subnetworks of the deceased patient. Disconnected individuals from the ego HCP are circled in brown.
}

\newcommand{\CaptionNIHCaseOneTwo}{
The development of reachable subnetworks over time for the survived patient's network chosen in \autoref{fig:ui}(a). The top and bottom rows depict the expanding reviewed and disseminated sets after each critical treatment stage, indicating that the selected HCP actively reviewed and wrote new EHR documents promptly.
}

\newcommand{\CaptionNIHCaseTwo}{
(a) compares a non-cyclic pattern with a cyclic pattern.
(b) and (c) present the distributions of the \textit{hierarchy} measure for the same two representative networks selected in \autoref{fig:ui}(a) and \autoref{fig:ui}(b).
(b1, b2, c1, c2) represent the reviewed and disseminated sets in local information flow diagrams, enabling a comparative study to identify underutilized EHR documents circled in brown.
}

\newcommand{\CaptionNIHCaseThree}{
The global information flow diagrams validate the influence from the distribution of HCPs' EHR reviewing accesses. By encoding node opacity and color with node-level \textit{in-closeness}, (a) illustrates unevenly distributed note-reading duties over HCPs in a deceased patient's treatment network, while (b) shows a case of more evenly distributed collaboration in a survived patient's treatment. This encoding helps identify two HCPs with high and low node-level \textit{in-closeness}, circled in purple and orange in (a), where their reviewed reachable subnetworks are shown in (a1) and (a2).
(a1) ensures that the HCP with high node-level \textit{in-closeness} had much less EHR reading responsibility compared to the HCP with low node-level \textit{in-closeness} in (a2).
}
\section{Introduction}
\label{sec:intro}
Healthcare professionals (HCPs) collaborate as a team in patient care, contributing at different treatment stages and locations. Electronic health records (EHRs) serve as a distributed communication system for asynchronous HCP communications, aiding in storing and retrieving patient information. 
HCPs collaboration is critical to the treatment quality and hence, can be associated with the patient's survival outcome \cite{gurses2006systematic, smits2010exploring, bagnasco2013identifying, verhaegh2017exploratory}.
Such collaboration around a patient may be examined as a time-evolving bipartite network, in which the nodes depict both the communicating HCPs and the accessed EHR documents, while edges signify information flows. 
To identify and foster successful teamwork~\cite{RN19,RN20,RN15}, assessing the network's \textit{effectiveness} involves understanding structural patterns that associate with better patient outcomes. Simultaneously, \textit{efficiency} in information flow speed and frequency must be evaluated.

Existing efforts in healthcare-related communication and EHR associations have different motivations from ours. 
In health informatics literature, some focused on analyzing HCPs' communication skills with patients \cite{butow2020using, pires2014communication, atkins2019assessing, de2020health, bagnasco2013identifying}, but little on teamwork evaluation. 
Some machine learning (ML) methods have been developed for representing relational information in EHRs \cite{wu2019representation, liu2020heterogeneous, cai2022hypergraph}. 
However, the generated representations are often challenging to interpret, and the information propagation within relations is insufficiently explored. 
More studies are needed to associate patients' survival outcomes with HCPs' teamwork structure and dynamics to enhance patient care quality.

Understanding structural patterns influencing communication outcomes allows researchers to intervene in future underperforming teamwork. 
Previous attempts focused on constructing communication effectiveness metrics from basic network measures. 
However, the evaluation criteria in these attempts are often limited and cannot be applied to assess other collaborations. 
In addition, some basic network measures quantify node-level structural features, but evaluating a network requires assessing network-level properties. Descriptive statistics, such as averages or maximum values, do not effectively aggregate certain node-level features into meaningful network-level representations. 
Addressing these challenges is crucial for developing communication effectiveness metrics.

Studying communication flow dynamics reveals delays in information propagation within a team. 
Network science researchers have developed temporal measures \cite{tang2009temporal, santoro2011time, kossinets2008structure, holme2005network} for analyzing network dynamics. Visualization designs~\cite{beck2017taxonomy, gleicher2011visual} have been introduced to convey information propagation over time. 
However, comparing among temporal measures can challenge human cognitive capability, as can interpreting information propagation from a single static visualization. To minimize perceptual burden, a new design for presenting perceivable communication efficiency is necessary.

In this work, we present a set of network analysis methods as contributions to address the outlined challenges. 
For assessing communication \textit{effectiveness}, we develop a flexible process to extract metrics that rate the success of teamwork structures, adaptable to diverse collaboration assessment criteria by exploring a broader range of network measures. Additionally, we formulate a method to aggregate node-level measures into network-level expressions based on distribution imbalance. 
To inspect communication \textit{efficiency}, we employ multiple views to assist users in configuring collaboration dynamics, simplifying the complexity of information propagation and temporal measures. 
To support the interpretation of communication \textit{effectiveness} and \textit{efficiency}, we design a visual analytics system that succinctly guides users in understanding the composition of effectiveness metrics and exploring the efficiency of information traveling through the network.
\section{Related Works}
\label{sec:relatedWorks}
In this section, we will discuss previous efforts in analyzing healthcare communication's influences and the assessment of general collaboration's effectiveness and efficiency. 
Current health informatics literature has a limited focus on associating patients' survival outcomes with HCPs' teamwork structure and dynamics. 
Regarding general communication assessment, though various methodologies have been applied to evaluate structure or dynamics in collaborations, there is a lack of concurrent examination of the two inter-related aspects in temporal network analysis.

\subsection{Health Informatics Analysis}
    To improve patient care quality, diverse health informatics analyses have been developed. 
    Previous works have linked treatment outcomes or patients' satisfaction acquired in survey format to HCPs' conversing protocols or skills analyzed with natural language processing (NLP) \cite{butow2020using, pires2014communication, atkins2019assessing, de2020health, bagnasco2013identifying}. 
    However, HCPs' teamwork evaluation stays at the debate level, where different HCPs have distinct perceptions of interprofessional collaborations \cite{verhaegh2017exploratory}.
    Some attempts utilize machine learning to encode complex relations among EHRs into representations \cite{wu2019representation, liu2020heterogeneous, cai2022hypergraph}, which can support downstream tasks such as disease diagnosis prediction. 
    Although these representations might generate accurate predictions, the relations associated with the predictions can hardly be understood and, hence, can barely assist in making meaningful interventions. 
    The limited research studying the collaboration among HCPs while producing interpretable analytical results needs to be resolved.

\subsection{Network Effectiveness Analysis}
    Network effectiveness analysis pertains to the investigation into better collaboration structure. 
    We observe an ubiquitous analysis workflow, consisting of (1) extracting topological characteristics using network measures, (2) uncovering distinguishing network measures through network comparison methods.

    \subsubsection{Network Measures}
    \label{sec:relatedWorks_measures}
        Network measures provide topological summaries in various aspects such as degree and betweenness centrality. 
        A full list of network measures and their definitions can be found in \cite{newman2018networks}. 
        Previous publications assessing cooperation effectiveness in different fields devote their efforts to examining a small set of measures. 
        To evaluate software developers' teamwork, Kidane and Gloor \cite{kidane2007correlating} leverage betweenness centrality and density, while Wolf et al. \cite{wolf2009predicting} additionally utilize degree centrality.
        Among healthcare literature, Creswick and Westbrook \cite{creswick2010social} review the medication advice-seeking interactions through density and the shortest path distance, whereas Zhu et al. \cite{zhu2019measuring} empirically observe the closeness and betweenness centrality distributions of HCPs' collaboration networks.
        However, these works explore only limited topological characteristics, and the evaluation criterion varies from one application to another.
        
    \subsubsection{Network Comparison}
    \label{sec:relatedWorks_comparison}
        Network comparison algorithms assist in discovering the similarities and differences in topological characteristics between opposite outcomes. 
        Aside from manually spotting \cite{wolf2009predicting, creswick2010social} the distinguishing factors, some studies enable automatic network comparison by utilizing random graph models or machine learning models. 
        Random graph model simulated networks construct a baseline distribution for each network measure \cite{dunn2011interpreting, latora2001efficient}. 
        Then, we locate each network measure extracted from a real-world network in the random graph baseline distribution, where the located percentile implies the uniqueness of this real-world network in terms of the inspected measure. 
        Yet, most simulated random graphs resemble real-world networks' topological characteristics within the locally tree-like structures \cite{callaway2000network}. 
        This premise might not be applicable for comparing global (graph-level) statistics (e.g., density). 
        In addition, random graphs are simulated specifically for establishing certain local nodal properties such as a scale-free degree distribution \cite{borgatti2009network, borgatti20092}. 
        Without diverse structural representations, these random graphs seem not to be a suitable baseline ensemble for our complex bipartite temporal networks. 
        Among emerging machine learning approaches to network comparison, Fujiwara et al. \cite{fujiwara2022network} utilizes contrastive learning \cite{zou2013contrastive}, which obtains network representations reflecting their distinct features. 
        Nonetheless, this work does not explore the combinational effect of structural features. 
        Lu et al. \cite{lu2023visual} presents a workflow to extract associations among nodal topological and semantic features of a multivariate network by constructing a composite variable that is the most correlated with the chosen target variable, enabling the investigation of the combinational influences. 
        Although this approach is potentially capable of composing effectiveness metrics for diverse scenarios, it is developed for nodal comparison in a static multivariate network.

\begin{figure}
    \centering
    \begin{subfigure}[b]{0.3\linewidth}
        \centering
        \includegraphics[width=\linewidth]{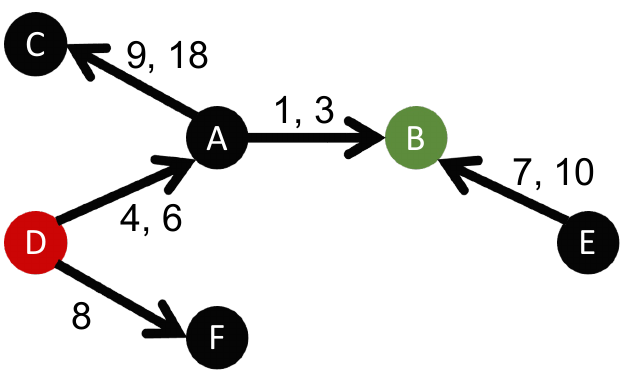}
        \caption{$t=0$. No events have occurred.}
        \label{fig:timepath_0}
    \end{subfigure}
    \hfill
    \begin{subfigure}[b]{0.3\linewidth}
        \centering
        \includegraphics[width=\linewidth]{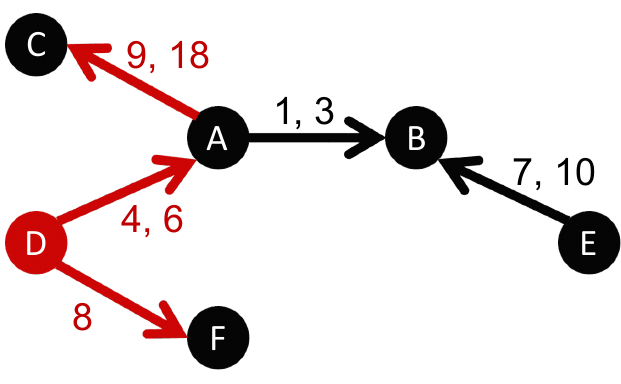}
        \caption{$t=\infty$. Influence set of node $D$, $[A, C, F]$.}
        \label{fig:timepath_infinite}
    \end{subfigure}
    \hfill
    \begin{subfigure}[b]{0.3\linewidth}
        \centering
        \includegraphics[width=\linewidth]{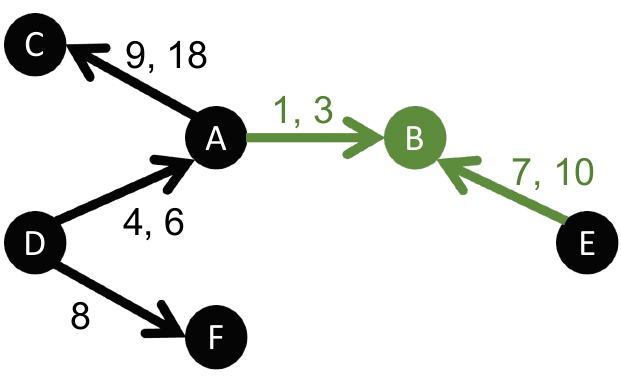}
        \caption{$t=\infty$. Source set of node $B$, $[A, E]$.}
        \label{fig:timepathsource_infinite}
    \end{subfigure}
    \caption{\CaptionTimePaths{}}
    \label{fig:timepaths}
\end{figure}

\subsection{Network Efficiency Analysis}
    Network efficiency analysis discovers and allows the inspection of dynamics in information flows.
    There are two main research directions, where one emphasizes temporal feature discovery and the other focuses on displaying the dynamics.

    \subsubsection{Temporal Measures}
    \label{sec:relatedTemporalMeasures}
        Compared to a static path, a \textit{time-respecting path} \cite{kempe2000connectivity} requires any two adjacent edges' timestamps happening in the same temporal order. 
        Tracing these time-respecting paths within a chosen observation time window, the \textit{influence set} of a chosen node $k$ is defined as the set of nodes that can be reached from $k$ through any time-respecting path, whereas the \textit{source set} of $k$ is the set of nodes that can reach $k$, as illustrated in \autoref{fig:timepaths}.
        The nodes in the influence set or source set construct a reachable subnetwork of $k$, and
        Holme \cite{holme2005network} further defines the \textit{reachability ratio} as the fraction of the number of nodes in the reachable subnetwork divided by the total number of nodes in the network.
        In addition, \textit{latency} measures the time it takes for a piece of information to be transmitted \cite{cooke1966shortest}, computed by the timestamps difference between the first and the last edge in a time-respecting path. 
        Transferring the strength measure from static networks, Kossinets et al. \cite{kossinets2008structure} calls this measure the \textit{frequency} of an interaction involving the same pair of entities.
        Tang et al. \cite{tang2009temporal} defines the \textit{temporal distance} between node $i$ and $j$ as the shortest time duration to form a time-respecting path between them and derive the \textit{global/local efficiency} measure as the reciprocal of the temporal distance.
        However, there are limited expeditious approaches for reviewing these abundant temporal measures.
        
    \subsubsection{Temporal Network Visualization}
        Temporal networks contain complex dynamics to be visualized. Designing representations for such data is not a trivial task.
        Previous designs emphasize displaying the evolution of network structure and can be sorted into a few main categories. We will discuss a simplified version of the taxonomy introduced in \cite{beck2017taxonomy}, focusing on the visualizations presented as node-link diagrams.
        \textit{Timeline} representations \cite{gleicher2011visual, reitz2010framework} present network structure snapshots along a timeline by stacking node-link diagrams or placing them side-by-side.
        \textit{1.5D timeline} approaches \cite{shi20141, liu2015egonetcloud} place a node's timeline at the center and connect its neighbor nodes to the timeline at the event occurring time (i.e., timestamp of the edge).
        \textit{Super-graph} visualizations \cite{loubier2008temporal, diehl2002graphs} include all existing time-respecting paths to provide a network overview. 
        Aside from designs presenting mainly topological evolution, \textit{TeCFlow} \cite{gloor2004tecflow} guides users step-by-step to narrow down the dynamics of interest to be investigated. 
        Yet, the guidance for each step is designed for a particular use case scenario and can hardly be generalized to other temporal networks.

\section{Analysis Goals and Tasks}
\label{sec:tasks}
As discussed in \autoref{sec:relatedWorks}, the existing literature has limitations in terms of (1) the generalizability of communication effectiveness evaluations, (2) the simplification of efficiency measure representations, and (3) the interpretability of both communication effectiveness and efficiency.
With our research goals being to tackle these three limitations, the corresponding analysis tasks are:
\begin{compactitem}
    \item \textbf{Task 1-1.} Narrow down each analyzed patient group with the similar demographics.
    \item \textbf{Task 1-2.} Enable flexible selection from the broad network measure search space.
    \item \textbf{Task 2-1.} Present each spatio-temporal feature in a separate visualization component.
    \item \textbf{Task 2-2.} Visualize temporal measures along with the evolution of network structure.
    \item \textbf{Task 3.} Empower concurrent explorations of interpretable effective structures and communication dynamics.
\end{compactitem}

\section{Data Processing and Analysis}
    In support of the analysis tasks outlined in \autoref{sec:tasks}, we have designed data processing and analysis methods, as well as a visual analytics system, to investigate the communication effectiveness and efficiency of HCPs.
    \subsection{Dataset Overview}
        The raw data comprises EHR digital traces of 518 patients diagnosed with Stage 2 or 3 breast, lung, and colorectal cancers, and the use of the data was approved by the IRB. 
        For each patient, the data contains his/her \textit{basic information} and the \textit{access logs} of their EHR data.
        The basic information includes each patient's demographics (e.g., age and sex), treatments, and survival outcome (alive/dead).
        EHR access logs include timestamped events from three months before to one year after the diagnosis date. 
        To maintain an equal collaboration timeframe, we exclude patients who passed away within a year after diagnosis.

        An EHR access event involves a HCP's document (either a note or a message) along with the HCP performing the writing (either creating or editing) or reviewing action. 
        Considering HCPs record a patient's medical conditions mostly in notes whereas messages contain information often without context, we only use the notes in this dataset. Hence, we use ``note'' and ``HCP/EHR document'' interchangeably throughout this paper. 
        In addition, we remove the administrative logs, such as billing events, which are not of interests when evaluating HCPs' teamwork influences toward the patients' survival. To specifically evaluate the core teams' communication, we include only the HCPs whose titles are MDs, RNs, ARNPs, PAs, Pharmacists, and Case Managers after consulting our medical doctor collaborators.
        
    \subsection{Network Construction}
        We construct a directed bipartite temporal communication network for each patient, encoding interactions between HCPs and EHRs based on access logs.
        The network contains two sets of nodes: the \textit{HCP nodes} and the \textit{note nodes} representing the HCPs and the EHR documents, respectively. 
        Since HCPs can review or edit a EHR document multiple times, there can be more than one timestamps recorded in each edge. 
        Therefore, a directed edge from a HCP node to a note node indicates that this core HCP created/edited this EHR document at the timestamp(s) recorded in this edge, whereas the other direction means that the HCP reviewed/accessed the note at the timestamp(s) recorded, fabricating the communications among HCPs through the EHR system.
        
    \subsection{Network Analysis of Communication Effectiveness}
        To extract the communication pattern effects on the patients' survival, we minimize the influences from the patients' demographics differences.
        Based on the advice from our medical doctor collaborators, we narrow down the analyzed patients using their cancer type, cancer stage, age, gender, and insurance (i.e., payer) before each run of the analyses (\textbf{Task 1-1}).
        
        \subsubsection{Deriving Network Measures through Node Aggregation}
        \label{sec:aggregation}
            To formulate a communication effectiveness metric for evaluating the HCPs' teamwork networks surrounding individual patients within a patient group, we provide a large metric search space consisting of network measures.
            We include the network-level measures and the node-level measures, which are aggregated into their network-level representations (\textbf{Task 1-2}).
            In this section, we introduce our selection of node-level measures and our node aggregation computations.
            The network-level measures in the metric search space will be discussed later in \autoref{sec:decomposition}.
            
            A node-level \textit{distance} 
            of a HCP node measures the average number of edges in the paths through which EHR documents are transmitted to this HCP. 
            We calculate the network-level \textit{distance} by averaging across all HCP nodes, revealing the overall connectivity of HCPs through message passing in the EHR system for a patient. 
            A node-level \textit{degree centrality} for a HCP node is defined as the number of edges connected to it. 
            However, common summary statistics such as the average or the maximum can hardly provide an overview of the \textit{degree centrality} of all nodes.
            The challenge persists when aggregating node-level \textit{betweenness centrality} and node-level \textit{closeness centrality}. In the former, a node's centrality is determined by the fraction of all-pairs shortest time-respecting paths that traverse through it, while in the latter, a node's centrality is gauged by the reciprocal of the sum of all-pairs shortest path distances (i.e., the number of edges in a path).

\begin{figure}
                \centering
                \includegraphics[width=0.50\linewidth]{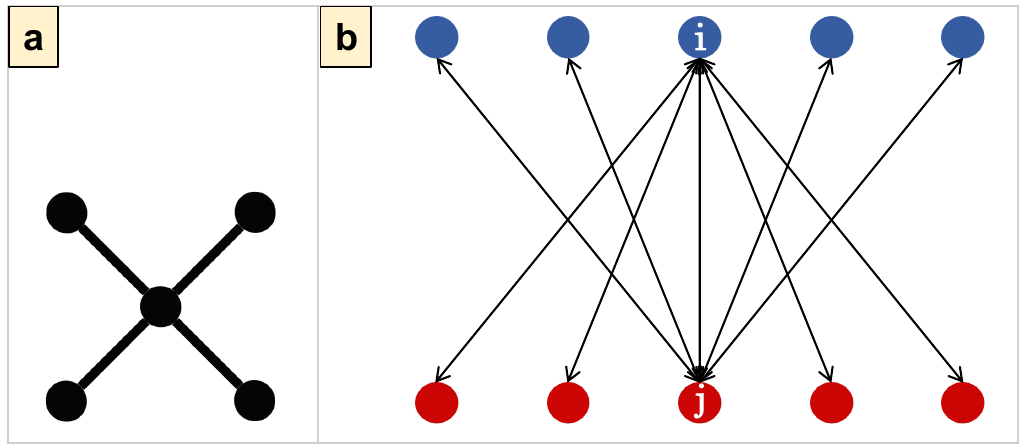}
                \caption{\CaptionStar{}}
                \label{fig:stars}
                \vspace{-0.1in}
            \end{figure}

            As a solution, Freeman \cite{freeman2002centrality} creates graph centralization index to reflect the extent to which the node-level measure distribution of all nodes in a network being imbalanced. 
            In other words, a higher graph centralization index indicates the distribution of a node centrality is being dominated by only a few nodes whose centrality values are much larger than the rest. 
            Freeman defined that for any node measure $X$ and a graph ${\cal G}$ with $n$ nodes, its graph centralization index is given by
            \begin{equation}
            \label{eqt:freeman}
                C_X({\cal G}) = 
                \frac{
                \sum_{k=1}^{n}\left(
                M_X(p^*,{\cal G}) - M_X(p_k,{\cal G})
                \right)
                }{
                \underset{{\cal G}^*}{argmax}
                \sum_{k=1}^{n}\left(
                M_X(p^*,{\cal G}^*) - M_X(p_k,{\cal G}^*)
                \right)
                } 
            \end{equation}
            , where $M_X(p_k,{\cal G})$ is the node measure value of node $p_k$, and $M_X(p^*,{\cal G})$ is the largest node measure value in the graph ${\cal G}$. The denominator is the maximum possible sum of node measure differences for any graph ${\cal G}^*$ with $n$ nodes. 
            The maximum possible sum of differences is proven by Freeman to occur in an undirected unipartite one-star topology as illustrated in \autoref{fig:stars}(a).
            
            We extend the graph centralization index definition to directed bipartite networks, applicable to our EHR communication networks
            in the appendix.
            Aggregated by our derivations of graph centralization index, the level of imbalance among various HCPs in the number of accesses to the EHR documents, in bridging communications, and in the distance to available EHR notes are separately indicated by the network-level \textit{degree}, the network-level \textit{betweenness}, and the network-level \textit{closeness}. 
            
        \subsubsection{Formulation of Effectiveness Metric}
            To enable a flexible selection from the broad metric search space, a general process customizing a collection of network measures for each patient group is required (\textbf{Task 1-2}). 
            Lu et al. \cite{lu2023visual} introduces a workflow for constructing a composite variable that is the most correlated with a chosen target variable from the nodal topological and semantic features of a multivariate network.
            We adapt this workflow into composing a communication effectiveness metric, which is the most associated with the patients' survival outcomes, from the network measure search space of all collaboration networks in the selected patient group.
            
            Specifically, we first train a four-layer multi-layer perceptron (MLP) to classify the patients' survival outcome. The MLP takes the network measure search space as input and produces the softmax probability of survival and death as output. After training the MLP, we proxy its output probability of survival using an optimized linear combination of input network measures.
            The set of weights assigned to the measures is the global optimal extracted through linear regression, that generates the largest possible Spearman's correlation coefficient between the probability of survival and the linear combination value.
            Subsequent to the optimization, we extract a threshold value which maximally separates the patients with the opposite outcomes by their linear combination values. 
            To be specific, we find the threshold which results in the largest multiple of the separation accuracy of each outcome.
            Finally, we have the linear combination values as the communication effectiveness metric scores, rating the potential success of HCPs' collaboration structures.
            In sum, we exploit the MLP's nonlinearity to discover complex associations and explain the discovered correlations through a semantically interpretable linear combination of network measures.
            By abstracting the communication effectiveness analysis into our series of optimizations, we enable the flexible selection of network measures by customizing linear combination weights for each patient group, representing the inter-relations among network measures quantifiably and interpretably.
            
        \subsubsection{Acquiring Node Importance through Network Perturbation}
        \label{sec:decomposition}
            On top of extracting communication effectiveness metrics, we facilitate detailed explorations to identify the importance of each HCP/note.
            To acquire such importance, we decompose network-level measures included in the metric search space to each node.
            
            We define this importance as the contribution of each node in maintaining the current value of a network-level measure. 
            If removing a node $k$ causes a significant change in a network-level measure, node $k$ contributes more to maintaining the current value than a node whose removal does not result in significant difference. 
            Therefore, we first perturb a network by removing a node, the edges connected to it, and the nodes that only connect to the removed node. Then, we re-compute each network-level measure in the metric search space. Finally, we calculate the absolute L1 distance between the original value of the network-level measure and the value after perturbation. 
            We repeat this decomposition process for each node in a network and for every network-level measure in the metric search space, including \textit{size} (i.e., the number of HCPs/notes), \textit{components} (i.e., the number of disconnected subnetworks within a network), \textit{density} (i.e., the ratio of the number of actual edges to the maximum possible edges of a network), and \textit{hierarchy} (i.e., the number of nodes that are not participating in any cyclic paths). 
    
    \subsection{Network Analysis of Communication Efficiency}
        Efficiency needs to be ingrained in the healthcare domain to disclose the 
        insights into the immediate accessibility and usage of disseminated notes among various HCPs and time periods,
        allowing the comparison between HCPs to identify the underperforming individuals.
        Addressing the timeliness of information dissemination, we extract the \textit{latency} of each EHR document access. 
        Concerning the perceived value of information, gauged by the number of times it has been utilized, we compute the \textit{frequency} of each note access.

        Unlike the \textit{latency} introduced in \autoref{sec:relatedTemporalMeasures}, we compute the \textit{latency} of each information access (i.e., each edge) by calculating the access timestamp difference between two adjacent edges in a time-respectin path.
        We define the first edge has zero latency as the EHR access occurs immediately. The second access has a react latency defined as the difference between the second edge's timestamp and the first edge's timestamp. The latencies for the rest edges in a time-respecting path are computed in the same way as the second edge.
        We repeat this computation for all time-respecting paths in the reachable subnetwork and then average the total \textit{latency} of an edge based on the number of times it participates in any time-respecting paths.
        
        Our \textit{latency} computation contrasts the reaction time of each EHR access and consequently records HCPs' relative speed in responding to updates in the EHR system.
        We compute the \textit{frequency} of each edge in the reachable subnetwork as the number of times of such event (i.e., reading or writing) involved in any time-respecting paths.
        Our calculation is grounded in the occurrences within distinct time-respecting paths. Each count signifies a distinct propagation, ensuring that an edge with earlier timestamp(s) does not receive preferential treatment over another with a larger count.

\begin{figure*}
    \centering
    \includegraphics[width=\linewidth]{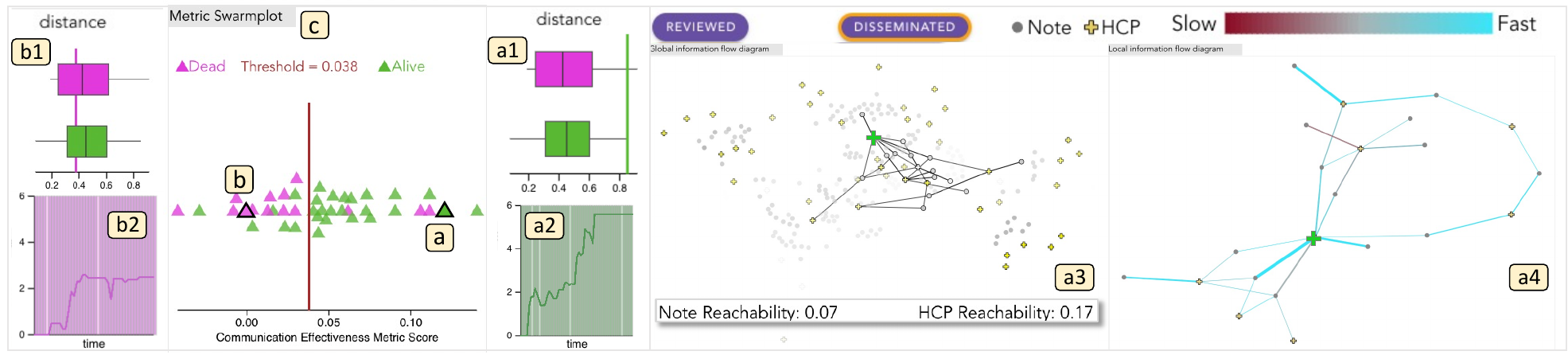}
    \caption{\CaptionUI{}}
    \label{fig:ui}
\end{figure*}
    
\section{Visual Analytics of EHRFlow}
    We design a set of visualizations, presenting distinct spatio-temporal features extracted by our network analyses (\textbf{Task 2-1}). 
    This design steers the user cognition to focus on apprehending one feature at a time, providing guidance for users to configure perceivable information flow diagrams that encode both temporal measures and the network structure evolution (\textbf{Task 2-2}).
    We integrate all supporting visualizations and configured diagrams into a visual analytics system, EHRFlow, empowering concurrent explorations of effective teamwork structures and communication dynamics (\textbf{Task 3}). 
    In this section, we walk through the analysis workflow steps embedded in EHRFlow and introduce each visual component.
       
    \textbf{Step 1: Effectiveness metric assessment.} In the \textit{Metric Swarmplot} as illustrated in \autoref{fig:ui}(c), each data point (triangle) is a patient's treatment network in the analyzed patient group and its color encodes the survival outcome. The communication effectiveness metric score of each patient is presented along the x-axis and the vertical positioning has no meaning but to prevent overlapping.
    Due to the complexity of temporal network characteristics, it is necessary to initially focus the inspection on a single communication network.
    Supported by the brown, vertical metric threshold line which
    separates the two outcomes, this visualization identifies the true positives (i.e., survived/alive), false positives, true negatives (i.e., not survived), and false negatives, 
    helping users configure a \textit{patient} (e.g., \autoref{fig:ui}(a) or \autoref{fig:ui}(b)) to examine their network's characteristics in detail.
    
    \textbf{Step 2: Network dynamics configuration.} 
    We present users with comprehensive metric statistics, including the ranks of network measures based on their weight magnitudes in the metric, as well as the metric correlation coefficient with the patients' survival, p value, and the separation accuracy using the metric threshold.
    The users can interpret the statistics as follows: 
    (1) a higher correlation value suggests that more distinctive topological characteristics exist between networks with contrasting outcomes;
    (2) a larger weight magnitude of a network measure represents a stronger association with the patients' survival;
    (3) a positive/negative weight means a larger value of this network measure potentially improves/deteriorates the chance of survival.
    
    Based on the metric composition statistics, 
    users can compare each network measure of the selected \textit{patient} to those of patients with opposite outcomes in the \textit{Network Measure Stacked Boxplot}.
    Notably, the network measure value we used in any stacked boxplot is network-level, which is either the value of a network-level measure or the aggregated value of a node-level measure.
    For example, \autoref{fig:ui}(a1) shows the aggregated distance measure stacked boxplot, displaying the survived (green) and the dead (pink) patients' aggregated distance measure distributions, consisting of their quantiles (i.e., 25th, 50th, and 75th percentiles) together with the minimum and the maximum at the two ends of the horizontal line.
    The green, vertical line locates the aggregated distance measure value of the \textit{patient} selected in \autoref{fig:ui}(a) on top of the stacked boxplot, colored with the \textit{patient}'s survival outcome and allowing users to conduct inter- and intra-outcome comparison.
    Another distance measure stacked boxplot example of the \textit{patient} selected in \autoref{fig:ui}(b) is presented in \autoref{fig:ui}(b1). 
    We provide stacked boxplots for all network measures in the overall interface, guiding users to decide a \textit{measure} of interests to configure the visual encodings of the information flow diagrams in \textbf{Step 3}.

    As Holme et al. \cite{holme2005network} states, the time window (\textit{timestamp}) needs to be defined before any investigation into a temporal network. 
    We enable the \textit{timestamp} configuration in the \textit{Network Measure Evolution Line Charts}.
    Same as the stacked boxplots, a line chart also displays either the value evolution of a network-level measure or the aggregated value evolution of a node-level measure.
    For instance, in \autoref{fig:ui}(a2), we provide the aggregated distance measure's evolution over the treatment duration of the \textit{patient} selected in \autoref{fig:ui}(a), where each bin is a 1-week interval in the \textit{patient}'s treatment and the bins colored with lighter green/pink contain critical treatment date(s) (e.g., chemotherapy). 
    Upon clicking a bin, all the bins before and included the clicked bin will be brushed to convey the analyzed time window (i.e., from the beginning to the clicked week in the treatment).
    Both \autoref{fig:ui}(a2) and \autoref{fig:ui}(b2) are the brushed line charts when the last bin is selected (i.e., to observe the entire treatment duration).
    We as well provide line charts for all network measures in the overall interface.
    
    \textbf{Step 3: Efficiency explorative evaluation.} 
    In \textit{Global Information Flow Diagram}, the users can explore the information propagation dynamics along with the network structural evolution. 
    The global diagram of the \textit{patient} selected in \autoref{fig:ui}(a) is illustrated in \autoref{fig:ui}(a3). We first draw the \textit{patient}'s network nodes whose locations are decided by the force-directed layout algorithm using every edge in time-respecting paths during  his/her recorded treatment. 
    The node opacity is decided by the selected \textit{measure} value, which is either the values of node-level measures or the decomposed values of network-level measures. Additionally, to enhance the visual contrast, note nodes are encoded by a white-gray color scale with circle shapes whereas HCP nodes are encoded by a white-yellow color scale with cross shapes based on also the value of selected \textit{measure}.
    Inspired by the \textit{1.5D timeline} design, which focuses on only an ego node's connections built over time, we enable the \textit{ego} configuration that can be selected based on the comparison among the \textit{measure} encoding of each node. 
    Upon selection, the \textit{ego} node's color is changed to pink or green according the \textit{patient}'s survival.
    
    To determine which edges to display in the global diagram, we use the single-source depth-first search (DFS) algorithm to retrieve time-respecting paths involving the chosen \textit{ego} node.
    Up until the selected \textit{timestamp}, the retrieved time-respecting paths form a reachable subnetwork of the \textit{ego} when clicking on either ``Reviewed'' (i.e., retrieve the source set) or ``Disseminated'' (i.e., retrieve the influence set).  
    Only the edges participating in this reachable subnetwork are visible in this view, and the reachable HCPs/notes' portions (i.e., reachability ratio) are computed and displayed along with their stroke being highlighted. 
    
    Finally, in \textit{Local Information Flow Diagram}, the users can explore how fast and how much information travels through the retrieved reachable subnetwork.
    The local diagram of the \textit{patient} selected in \autoref{fig:ui}(a) is illustrated in \autoref{fig:ui}(a4). We utilize the force-directed algorithm to redraw the \textit{ego}'s reachable subnetwork, depicting more explicit local structures. 
    The edge thickness encodes the information exchange frequency and the edge color encodes the reciprocal of latency, which is an HCP's react speed to an existing EHR document encoded by a maroon-gray-cyan color scale, mapped from slow to fast. 

\section{Case Studies}
    \subsection{Overview of Patient Groups and Analysis Goals}
        \textbf{Patient group A} consists of 41 patients ranging from age 65 to 75 with Stage 3 lung cancer, where 27 patients survived and 14 patients did not. 
        The correlation between teamwork topological features and these patients' survival is \texttt{0.5397} with the accuracy \texttt{0.7778} for the survived patients and \texttt{0.7857} for the deceased patients.
        We conducted Study 1 and Study 2 on this group, analyzing collaboration patterns separately to elucidate both disconnected communications and underutilized EHR documents.
        \vspace{-0.1in}
        \begin{table}[h]
            \centering
            \begin{tabular}{ |p{1cm}|p{1.3cm}|p{0.8cm}|p{1cm}|p{1cm}|p{1cm}| } 
              \hline
                Rank\#2 & \#4 & \#6 & \#7 & \#9 & \#11\\ 
              \hline
                Distance & Betweenness & Note Size & 
                Hierarchy & In-Closeness & Out-Degree\\
              \hline
                0.3868 & -0.3379 & 0.1999 & -0.1738 & -0.1279 & -0.058\\
              \hline
            \end{tabular}
            \vspace{0.02in}
            \caption{\CaptionNIHDataAMetric{}}
            \label{tbl:dataA}
            \vspace{-0.15in}
        \end{table}
            
        \textbf{Patient group B} contains 45 female patients ranging from age 65 to 75 with Stage 2 breast cancer, where 39 patients survived and 6 patients did not. 
        The collaboration structural features are strongly correlated with the patients' survival, having \texttt{0.9618} Spearman's correlation coefficient and the separation accuracy \texttt{0.8718} for the survived patients and \texttt{1.0} for the deceased patients using the metric threshold.
        We conducted Study 3 on this group, investigating the influences from the distributions of notes reading and writing duties among HCPs.
        \vspace{-0.1in}
        \begin{table}[h]
            \centering
            \begin{tabular}{ |p{1cm}|p{1.3cm}|p{1cm}|p{0.9cm}|p{0.8cm}|p{1cm}| } 
              \hline
                Rank\#1 & \#4 & \#6 & \#8 & \#10 & \#11\\ 
              \hline
                In-Closeness & Betweenness & Out-Closeness & Density & HCP size & Hierarchy\\ 
              \hline
                -0.4986 & -0.3094 & 0.2899 & -0.2637 & -0.119 & 0.1149\\
              \hline
            \end{tabular}
            \vspace{0.02in}
            \caption{\CaptionNIHDataBMetric{}}
            \label{tbl:dataB}
            \vspace{-0.15in}
        \end{table}
    
    \begin{figure*}
        \centering
        \includegraphics[width=0.7\linewidth]{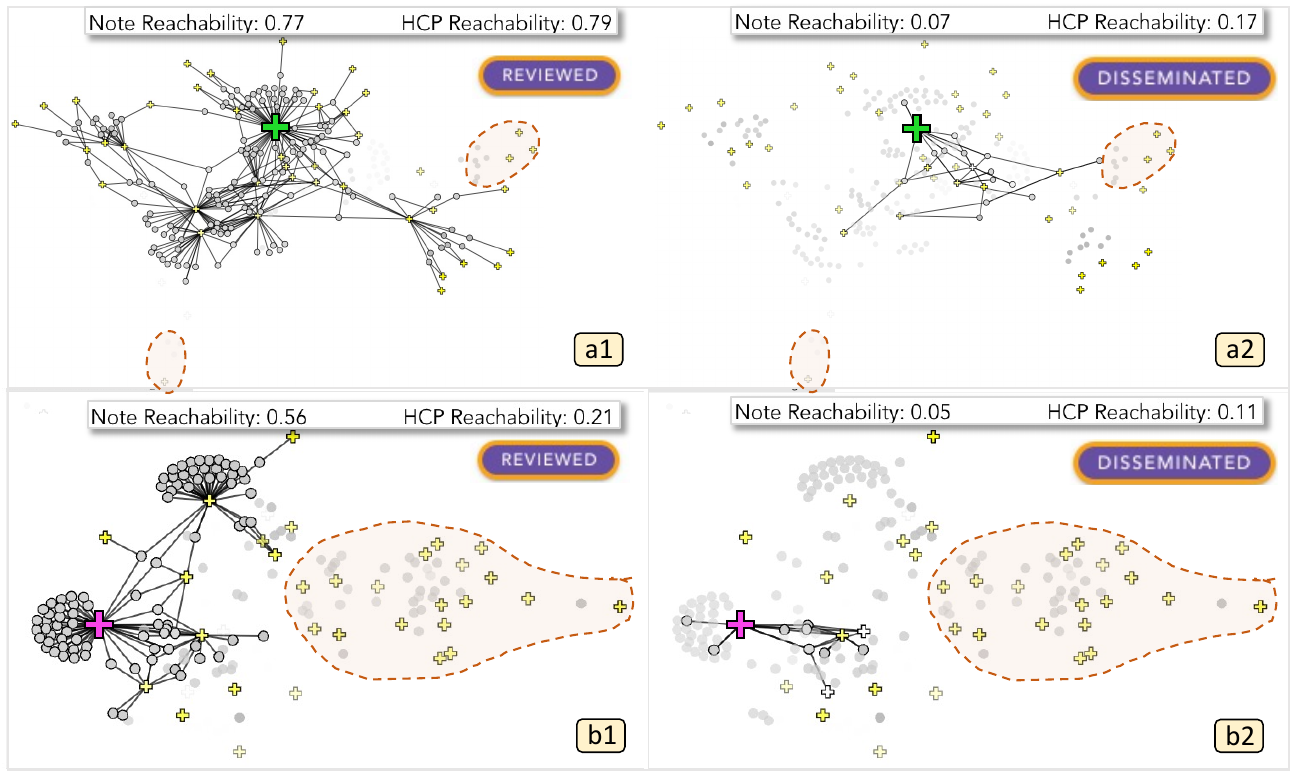}
        \caption{\CaptionNIHCaseOne{}}
        \label{fig:case1}
        \vspace{-0.1in}
    \end{figure*}
    \begin{figure}
        \centering
        \includegraphics[width=\linewidth]{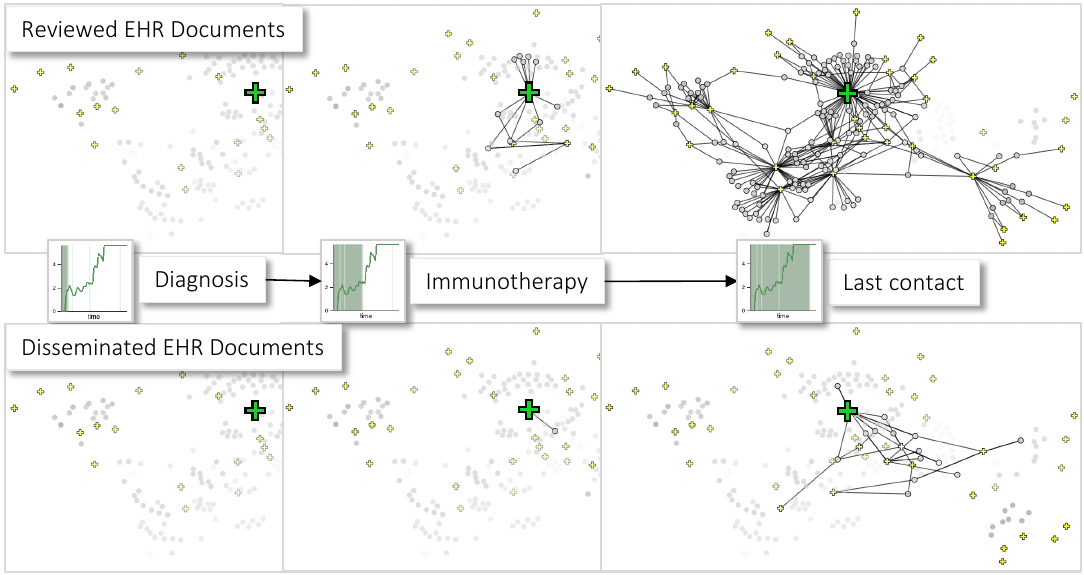}
        \caption{\CaptionNIHCaseOneTwo{}}
        \vspace{-0.1in}
        \label{fig:case1_2}
    \end{figure}
    \subsection{Study 1: Identification of Disconnected Communications}  
        In this study on the patient group A, we reveal potential information loss among HCPs characterized by a lack of internal communication throughout the treatment, as determined by the \texttt{distance} measure and the global information flow diagram.
        
        To begin with, we observe that \texttt{distance} is weighted \texttt{0.3868} in \autoref{tbl:dataA} and ranked as the second most associated network measure with the patient survival.
        As introduced in \autoref{sec:aggregation}, each node-level measure, such as \texttt{distance}, is aggregated to a network-level representation. 
        The aggregated \texttt{distance} represents the average number of edges traversed by information before reaching an HCP. 
        In our context, a larger aggregated \texttt{distance} value indicates that more HCPs are connected and involved in transmitting EHR documents, broadening the information accessibility. 
        Suggested by the positive weight of aggregated \texttt{distance} in the metric, the more HCPs involving in passing information around within a patient's treatment network, the more likely that this patient would have a better outcome. 
        To investigate and understand the potential causes,
        we select two representative patients (\autoref{fig:ui}(a, b)) having the opposite outcomes and relatively high and low aggregated \texttt{distance} values in \autoref{fig:ui}(a1, b1), and choose to examine their entire treatment duration in \autoref{fig:ui}(a2, b2).
        
        In \autoref{fig:case1}(a1, a2, b1, b2), we observe disconnected communications in the ego HCPs' reachable subnetworks.
        The two ego HCP nodes (i.e., the green cross and the pink cross) have close node-level \texttt{distance} values to the two networks' aggregated \texttt{distance} values, indicated by the vertical lines in their \texttt{distance} boxplots as shown in \autoref{fig:ui}(a1, b1). 
        Hence, the comparison between these two ego HCP nodes is representative of contrasting the two networks.
        In \autoref{fig:case1}(a1), the survived patient's reviewed reachable subnetwork shows the HCPs in the brown circle did not transmit any information to the ego HCP, and in \autoref{fig:case1}(a2), the disseminated reachable subnetwork presents the EHR documents created/edited by the ego HCP was never propagated to the HCPs in the brown circle.
        In comparison to \autoref{fig:case1}(b1, b2), a much larger disconnected portion is spotted in the brown circles within the deceased patient's treatment collaboration. 
        Quantifiably speaking, by comparing the reachability ratios in \autoref{fig:case1}(a1, b1), we find the ego HCP treating the deceased patient had three times lower reachability to other HCPs in the reviewed sets, and half as much reachability to both HCPs and notes when contrasting the reachable ratios of the two disseminated sets in \autoref{fig:case1}(a2, b2).
        Had more HCPs participated in passing on the information, the aggregated \texttt{distance} would have been larger and the information could have reached to these disconnected regions, potentially improving the treatment outcome.
        
        To confirm the growth in aggregated \texttt{distance} indeed helped HCPs reaching to more others, we choose three timestamps, where the collaboration networks exhibited progressively increasing values of aggregated \texttt{distance}, and explore the corresponding development of reachable subnetworks. 
        As evident in \autoref{fig:case1_2}, the reviewed (top row) and the disseminated (bottom row) reachable subnetwork kept expanding from the diagnosis date, through the start date of the immunotherapy, to the last contact date, indicating this ego HCP treating the survived patient actively accessed more information and created more notes after each stage of treatment.
        Thus, the association between the aggregated \texttt{distance} and information reachability is validated.
        
        By promoting desired communication patterns regarding the \texttt{distance} measure in a network, the overall effectiveness metric score of a patient can be improved toward the surviving end.
        While more in-depth experiments are needed, we summarize our findings as potential directions of improvements. 
        First, HCPs are encouraged to maintain a message passing structure that connects diverse HCPs to share EHR documents and increase the overall information accessibility. 
        Second, HCPs are prompted to access more available notes and create more EHR documents to be passed on, preventing the disconnected communications.
        Finally, the amount of reachable HCPs and notes should increase after each treatment stage, developing a timely teamwork.
        Through our network evolution line chart and global information flow diagram, users can monitor the development of the disconnected portion and intervene potential information loss in time before a flawed treatment decision could be made.
        
    \begin{figure}
        \centering
        \includegraphics[width=\linewidth]{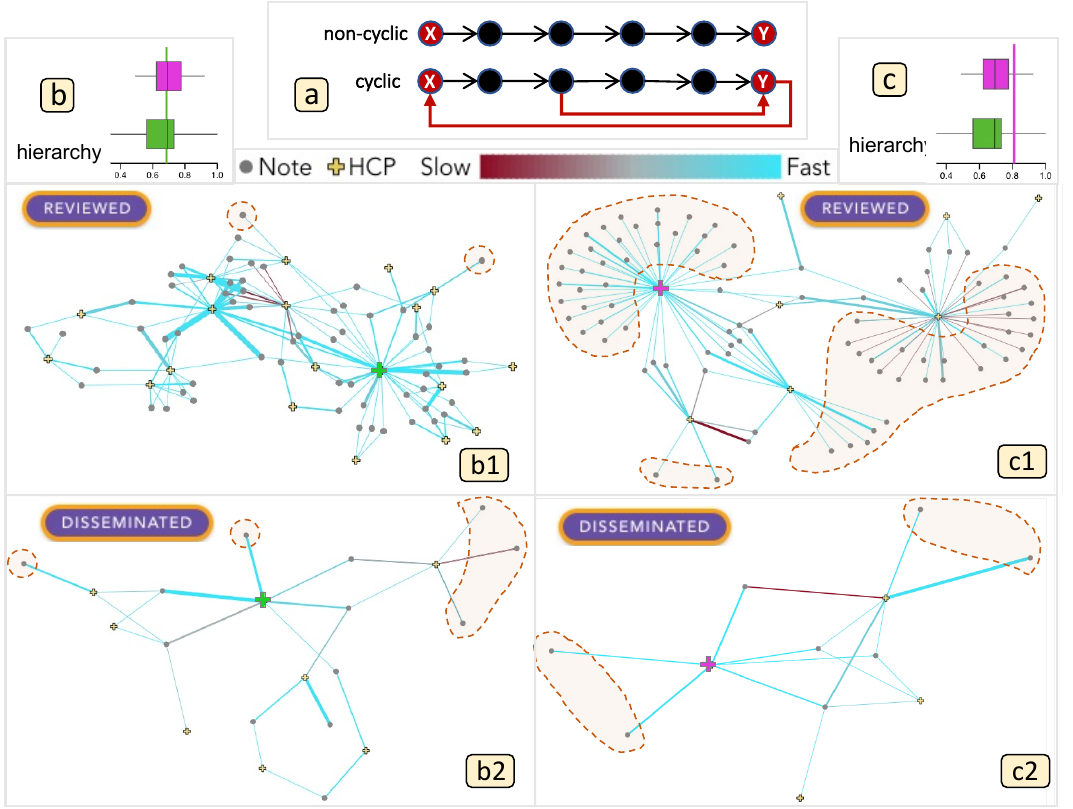}
        \caption{\CaptionNIHCaseTwo{}}
        \label{fig:case2}
        \vspace{-0.1in}
    \end{figure}
    \subsection{Study 2: Trade-Off of Notes Underutilization}
        Following Study 1, our continued investigation of the same two patients and ego HCPs in the patient group A uncovers the underutilization of EHR documents along with a trade-off between notes underutilization and disconnected communications, as indicated by the \texttt{hierarchy} measure and the local information flow diagram.
        
        To start with, 
        we observe that \texttt{hierarchy} is weighted \texttt{-0.1738} in \autoref{tbl:dataA} and only ranked 7th among all twelve measures.
        \texttt{Hierarchy} is defined as the count of nodes that are not part of any cyclic paths. In the context of our study, a lower \texttt{hierarchy} value suggests that more HCPs maintained a record of the information they transmitted, accessing EHR documents from HCPs who had reviewed the transmitted notes.
        Informed by \texttt{hierarchy}'s negative weight, more HCPs tracking their previously accessed notes correlates with higher patient survival chances.
        Drilling down to the local information flow diagrams, we discover the association between the \texttt{hierarchy} measure and the amount of underutilized EHR documents.
        In \autoref{fig:case2}(b1, c1), the brown circles contain root nodes that are the source notes of information propagation paths but were only accessed by one HCP in each reviewed reachable subnetwork. 
        In \autoref{fig:case2}(b2, c2), the brown circles highlight the leaf nodes (i.e., destination nodes of paths) that were never accessed by any HCPs in the treatment network.
        The root and leaf nodes signify messages transmitted but inadequately tracked by HCPs, stemming from non-cyclic patterns. Their quantity rises with an increase in the \texttt{hierarchy} value. 
        Contrasting the reachable subnetworks, the majority of nodes in the survived patient's network participates in cyclic paths (\autoref{fig:case2}(b1, b2)), and yet, the brown circled region containing the underutilized root/leaf nodes are much bigger in the treatment network of the chosen deceased patient (\autoref{fig:case2}(c1, c2)).

        The above observation suggests a preference for a lower \texttt{hierarchy}; however, a reduction in \texttt{distance} is linked to a decrease in \texttt{hierarchy}. As depicted in \autoref{fig:case2}(a), the additional two red-colored edges in the cyclic pattern shortened the \texttt{distance} from node $X$ to node $Y$ compared to the non-cyclic pattern. 
        Consequently, we aim to interpret the potential trade-off between these two measures.
        As seen in \autoref{fig:ui}(a1) and \autoref{fig:case2}(b), the patient who survived had a relatively higher \texttt{distance} and lower \texttt{hierarchy}, while the deceased patient showed the exact opposite, as depicted in \autoref{fig:ui}(b1) and \autoref{fig:case2}(c). These divergent patterns affirm that the two sampled patients are fitting representatives for exploring the trade-off.
        Despite a shorter \texttt{distance} not being a desirable teamwork feature, as outlined in Study 1, the metric still assigns a negative weight to \texttt{hierarchy}, prioritizing the importance of minimizing underutilized notes.
        Conversely, placing greater emphasis on the magnitude of \texttt{distance} underscores the continued significance of preserving the message passing structure that links various HCPs. 
        This leads to an optimal teamwork pattern, ensuring effective tracking of notes among closely communicated HCPs while also disseminating them to other HCP groups.
        
        In our local information flow diagrams, users can intuitively identify HCPs with slower reaction speeds to EHR documents based on edge colors. Additionally, they can pinpoint underutilized EHR documents by leveraging the explicit local structure. This implies a potential opportunity for intervention to minimize wasted efforts in creating notes.
        
    \begin{figure*}
        \centering
        \includegraphics[width=\linewidth]{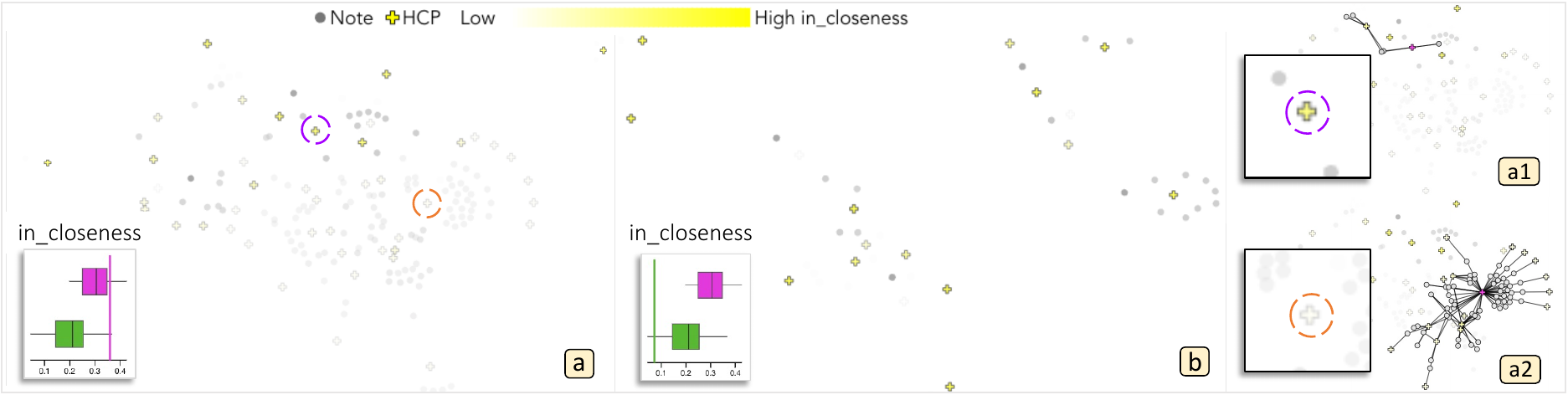}
        \caption{\CaptionNIHCaseThree{}}
        \label{fig:case3}
        \vspace{-0.1in}
    \end{figure*}
    \subsection{Study 3: Influences from EHR Reading and Writing Duty}
        In this investigation of patient group B, we examine the distinct influences from network measures that signify the varying levels of responsibility among HCPs in reading and writing EHR documents. This examination is informed by the global information flow diagram and the measures of \texttt{in-closeness} and \texttt{out-closeness}.
        
        In \autoref{tbl:dataB}, we observe that the aggregated \texttt{out-closeness} has the opposite weight sign to the aggregated \texttt{in-closeness}. As a recap of our discussion in \autoref{sec:aggregation}, node-level \texttt{in-closeness} and \texttt{out-closeness} represent the reciprocal of the sum of all-pairs shortest path distances flowing into and out of a node, respectively. Therefore, these measures reflect the inverse of the information influx and outflux to and from an HCP, indicating the HCP's EHR reading (influx) and writing (outflux) loads.
        Aggregated through our graph centralization index derivations, the network-level values of these two measures uncover the extent of imbalance among different HCPs in the aforementioned node-level semantics. Consequently, the positive weight of aggregated \texttt{out-closeness} suggests a positive correlation between patients' survival and the imbalanced distribution of node-level \texttt{out-closeness} within patient group B. In simpler terms, regarding the EHR writing duty of every HCP within a treatment network, the more imbalanced the distribution, the more likely the collaboration leads to a better survival outcome.
        On the contrary, with the negative weight of \texttt{in-closeness}, a more unevenly distributed EHR reviewing access over every HCP tends to result in a less favorable chance of survival.

        Given the aggregated \texttt{in-closeness} is considered twice as much associated with the patients' survival as the aggregated \texttt{out-closeness} according to \autoref{tbl:dataB}, we validate our interpretations by inspecting two representative treatment networks with the opposite outcomes. 
        In \autoref{fig:case3}(a), this sampled deceased patient has a large aggregated \texttt{in-closeness} value (i.e., the vertical line on top of the stacked boxplot) among all patients (i.e., the stacked boxplot distributions), indicating the EHR reading responsibility was unevenly distributed among HCPs collaborated in this treatment. 
        In the global information flow diagram of \autoref{fig:case3}(a), this imbalanced distribution among HCP nodes is displayed by having distinctively node encodings, showing high (i.e., opaque yellow) and low (i.e., transparent white) node-level \texttt{in-closeness} values.
        In comparison to the sampled survived patient whose network exhibits a lower aggregated \texttt{in-closeness} as indicated in the stacked boxplot of \autoref{fig:case3}(b), most HCP nodes have high node-level \texttt{in-closeness} values, showing a more preferable even distribution of EHR reading assesses suggested by the metric. 
        The encoding differences further suggests us to investigate the individual HCPs' EHR reading responsibility, where we expect to observe HCPs with higher node-level \texttt{in-closeness} (i.e., the inverse of the information influx) having lower accesses in reading notes.
        As validated in \autoref{fig:case3}(a1), the reviewed reachable subnetwork of a representative ego HCP who has larger node-level \texttt{in-closeness} is much smaller than the one of an ego HCP who has lower \texttt{in-closeness} as shown in \autoref{fig:case3}(a2).

        In sum, our metric uncovers a favorable collaboration pattern, centralizing the EHR writing responsibility on a few HCPs, but requiring every HCP to have similar diligence in reading notes.
        The global information flow diagram validates this observation by showing a clear overview of node-level \texttt{in-closeness} using node encodings, also enabling the identification of HCPs being loaded with high/low reading responsibilities and providing clues for potential mediation in reassigning EHR duties.

\section{Expert Feedback}
    \label{sec:expertFeedback}
    To evaluate our analysis system's usability, besides the extensive case studies, we conducted an interview of experts in healthcare domain. 
    The first expert (E1) is a medical doctor who is also on the  faculty of internal medicine.
    Both the second expert (E2) and the third expert (E3) are 
    on the faculty of Public Health.
    The interview was conducted through a video conference setting, where the analysis methodology was explained and the  
    three case studies and corresponding insights were presented via real-time interactions with the EHRFlow interface.

    Regarding the data usage, E1 raised a concern, ``Patient's comorbidity is often considered to have significant association with the patient's survival. But this information is not included in the analysis.''
    Though patient's comorbidity can be inter-related to the teamwork structure and the patient's survival, we find the influences on the survival outcome from the comorbidity and from the collaboration structure should be inspected sequentially or separately.
    We provide evident justifications of this claim using the following two scenarios.
    First, the comorbidities of the analyzed patient group A have \texttt{0.866} Spearman's correlation with the patients' survival whereas the effectiveness metric scores have \texttt{0.5397}. 
    Our effectiveness metric score extracts interpretable collaboration patterns adopted by HCPs to treat the patients with worse comorbidities, where researchers can potentially investigate if there were structural interventions that could have been made to improve the treatment quality. 
    Second, the comorbidities of the analyzed patient group B have \texttt{0.6223} Spearman's correlation with the patients' survival but the effectiveness metric scores have \texttt{0.9618}. 
    In this case, researchers have more opportunity to further investigate the excellent teamwork that helped the patients having worse comorbidities survive, and promote the communication patterns in the future treatment of this patient group. 
    The two examples validates the separate correlations (i.e., highly different correlation coefficients) from the comorbidity and from the communication structure.
    Notably, there are many popular comorbidity indices such as the Charlson Comorbidity Index \cite{charlson1987new} and the Elixhauser Comorbidity Index \cite{elixhauser1998comorbidity}. 
    To prevent the debate of method selection here, we did not adopt these approaches but used linear regression to find the optimal correlation between the comorbidity and the survival outcome. 
    In the future, we plan to consult healthcare experts to incorporate the patient's comorbidity with domain considerations into our analysis workflow.

    E2 considered our analysis design is proper because our analysis method always takes into
    account the period before the patient was diagnosed with cancer. 
    E3 questioned ``As your method aims to analyze the associations with patient survival, why don't you utilize survival analysis?''  
    While survival analysis can predict the time to an event (death or cure), our primary goal is to unveil teamwork patterns associated with patients' survival. This involves contrasting around 40 historical patients' EHR-mediated communication networks with known outcomes. Extracted trends offer insights into potential patterns for future patients. Domain experts should carefully inspect individual patients before considering interventions based on our suggested insights. Furthermore, we have filtered patient demographics for a specific group and ensured an identical timeframe for deceased and survived patients, eliminating sickness level bias in the extracted associations.
    E1 also noted, ``As healthcare professionals, we need to fully understand the meaning of each network measure to utilize EHRFlow.''  Admittedly, to use EHRFlow, certain extent of learning is required. To assist healthcare professional users in applying the network measures, we plan to provide intuitive interpretations through the user interface.

    There were several appraises of our visual analytics interface. 
    E1 commented, ``It is valuable that this visual analytics interface enables us to interactively drill down to identify the high/low value EHR documents and to observe the information dissemination (i.e., over a reachable subnetwork).''
    E1 added, ``The drilling down investigation process exposes the potential interventions that could have been carried out for identified individuals.''
\section{Discussion and Limitations}
    Through the case studies and the expert feedback, we demonstrate the capability and usability of EHRFlow.  
    Besides the limitations and justifications of our methods discussed in \autoref{sec:expertFeedback}, we consider  additional alternative approaches here.
    
    \textbf{Alternative for extracting effectiveness.} 
    A popular alternative is to train a graph neural network (GNN) to classify the network outcomes based on the extracted network representations and generate explanations of the GNN's behavior.
    While GNNs might outperform MLP in extracting associations for network data, their behavior is much harder to be interpreted.
    Existing GNN explanation works \cite{ying2019gnnexplainer, vu2020pgm} often generate a subnetwork as the most important substructure contributing to the classification results, which can still be quite complex and hardly be further reasoned that why this structure is undesirable when ground truth is unavailable.
    Hence, a more understandable type of GNN explanation is required and we plan to investigate this direction in future research. 
    
    \textbf{Alternative for exploring efficiency.}
    One alternative is to employ the global and local efficiency measure derived in \cite{tang2009temporal, latora2001efficient}, which separately computes the average speed in information accessing over the entire network or over the edges connecting to a node's 1-hop neighbors.
    However, the global efficiency measure does not support identifying individual HCPs with lower efficiency and the local efficiency includes unwanted dynamics from averaging over the edges outside the reachable subnetwork.

    \textbf{Additional node attributes.}
    We also plan to incorporate HCPs' information (e.g., primary doctors or specialists) other than their titles.
    Knowing more about the investigated HCP nodes or even the content of examined EHR document nodes during the exploration potentially helps derive deeper insights. 
    We plan to provide this information after the proper deidentification of data.
\section{Conclusion}
    We introduce a novel approach to leveraging EHR data for potentially enhancing healthcare outcomes. Our resulting visual analytics system, EHRFlow, enables the assessment of communication effectiveness and efficiency among healthcare professionals by analyzing and visualizing communication networks constructed from EHR access logs. 
    Our case studies showcase the potential capability and value of EHRFlow, particularly with cancer patient datasets. We verified our analysis results and gathered valuable feedback for further research through reviews by healthcare experts. While our study has primarily focused on healthcare teamwork using EHR data, the design of EHRFlow may find applications in the analysis of other teamwork scenarios, spanning from software development projects and scientific research collaborations to disaster management and beyond.

\bibliographystyle{IEEEtran}
\bibliography{00_ref}

\appendix
\vspace{-0.07in}
To aggregate node-level measures, where summary statistics do not provide meaningful representations of all nodes, we extend the graph centralization index \cite{freeman2002centrality} definition to directed bipartite networks, which is applicable to our EHR communication networks.
Inheriting the core concepts from Freeman's equation \autoref{eqt:freeman}, we derive that a two-star topology as illustrated in \autoref{fig:stars}(b) leads to the maximum possible sum of differences for degree, betweenness, and closeness centrality in the following paragraphs.

The two-star topology includes a star node $i$ in the node set $V_1$ containing $n_1$ nodes, bi-directionally connected to all nodes in the other node set $V_2$, along with another star node $j$ in $V_2$ containing $n_2$ nodes, bi-directionally connected to all nodes in $V_1$. 
Using these notations, we will derive the equations for directed bipartite graph centralization indices in the following paragraphs. 

\textit{Degree centrality} for a node $k$ is the number of edges connected to $k$, which can either be \textit{in-degree} or \textit{out-degree}, counting the incoming or outgoing edges, respectively.
Taking \textit{out-degree} as example, for the star node $i$, the centrality difference from a non-star node in $V_2$ is $(n_2-1)$ and there are $(n_2-1)$ non-star nodes in $V_2$, which leads to $(n_2-1)^2$; the centrality difference from a non-star node in $V_1$ is $(n_2-1)$ and there are $(n_1-1)$ non-star nodes in $V_1$, which results in $((n_2-1)*(n_1-1))$; the centrality difference from the star node $j$ is $(n_2-n_1)$.
Finally, we have the maximum sum of degree centrality differences from the star node $i$ in $V_1$ as the sum of the above three terms $Degree_{diff\_V_1} = (n_2-1)^2 + (n_2-1)*(n_1-1) + (n_2-n_1)$.
Deriving the same for the star node $j$, $Degree_{diff\_V_2}$ can be computed and the directed bipartite degree graph centralization index $C_D$ can be obtained by:
\begin{math}
    \resizebox{3.5in}{!}{$
    C_D({\cal G}) = \frac{\sum_{k=1}^{n_1} [M_D(p^{*}_{V_1},{\cal G})-M_D(p^{ }_{V_1k},{\cal G})]}{Degree_{diff\_V_1}} 
     + \frac{\sum_{k=1}^{n_2} [M_D(p^{*}_{V_2},{\cal G})-M_D(p^{ }_{V_2k},{\cal G})]}{Degree_{diff\_V_2}}$}
\end{math}
, where $M_D(p^{ }_{V_1k},{\cal G})$ can be either the in-degree or out-degree centrality value of node $p^{ }_{V_1k}$ in the node set $V_1$, and so can $M_D(p^{ }_{V_2k},{\cal G})$.

The star structures induce the maximum sum of centrality differences because they maximally centralize the minimum number of node(s).
We further validate the usability of our two-star topology by analyzing the potential edge rewiring.
Within the two-star directed bipartite topology, the only two rewiring options are (1) removing the edge(s) between star nodes $i$ and $j$, and (2) rewiring edge(s) to connect two non-star nodes in the different node sets.
First, we recompute the three out-degree difference terms in $Degree_{diff\_V_1}$ after removing the edge from the star node $i$ to the star node $j$, which will be $(n_2-2)*(n_2-1)$, $(n_2-2)*(n_1-1)$, and $(n_2-n_1-1)$. The total centrality difference after the edge removal is $(n_2+n_1-1)$ less than before.
Second, after rewiring the source node of an edge from the star node $i$ to a non-star node in $V_1$, the same three terms of out-degree centrality difference will be $(n_2-2)*(n_2-1)$, $(n_2-2)*(n_1-2)+(n_2-3)$, and $(n_2-n_1-1)$. The total centrality difference after edge rewiring is $(n_2+n_1)$ less than before. 
Therefore, the two-star topology remains the structure resulting in the maximum sum of centrality differences.

\textit{Betweenness centrality} for a node $k$ is the fraction of all-pairs shortest time-respecting paths that pass through node $k$.
We adopt the modified normalization term defined in \cite{borgatti2011analyzing} to compute directed bipartite betweenness centrality $M_{B\_V_1}$ and $M_{B\_V_2}$. 
In addition, the denominator of betweenness graph centralization index was defined in \cite{freeman1977set} as the number of nodes, which will be $n_1-1$ and $n_2-1$ in bipartite networks. Hence, we derive the directed bipartite betweenness graph centralization index $C_B$ as:
\begin{math}
    \resizebox{3.5in}{!}{$
    C_B({\cal G}) = \frac{\sum_{k=1}^{n_1} [M_{B\_V_1}(p^{*}_{V_1},{\cal G})-M_{B\_V_1}(p^{ }_{V_1k},{\cal G})]}{n_1-1}
     + \frac{\sum_{k=1}^{n_2} [M_{B\_V_2}(p^{*}_{V_2},{\cal G})-M_{B\_V_2}(p^{ }_{V_2k},{\cal G})]}{n_2-1}.$}
\end{math}

\textit{Closeness centrality} of a node $k$ is the reciprocal of the sum of all-pairs shortest path distance (i.e., the number of hops) of $k$ to(out-closeness) or from(in-closeness) all the other nodes, normalized by the minimum possible sum of distances \cite{freeman2002centrality, wasserman1994social}. 
For bipartite networks, the minimum distances in the two node sets are $(2 * (n_1 - 1) + n_2)$ and $(2 * (n_2 - 1) + n_1)$ as discussed in \cite{borgatti2011analyzing}, which can be resulted from the two star nodes $i$ and $j$, respectively.
% This derivation again validates that the two-star topology induces the 
This observation further validates that our two-star topology optimally centralizes the minimum number of nodes, and from which we can derive the closeness node measure for separate node sets as $M_{C\_V_1} = \frac{2 * (n_1 - 1) + n_2}{d_{V_1}}$ and $M_{C\_V_2} = \frac{2 * (n_2 - 1) + n_1}{d_{V_2}}$, 
where $d_{V_1}$ and $d_{V_2}$ are the sum of all-pairs shortest path distance.

Since the normalized terms in $M_{C\_V_1}$ and $M_{C\_V_2}$ are constants w.r.t any edge rewiring, we analyze the potential minimum shortest path distance to find the largest possible closeness measure difference in a directed bipartite network.
In our two-star topology, the closeness node measure is $1$ of the star nodes $i$ and $j$. For a non-star node $k$ in $V_1$, the shortest path distance from another node in $V_1$ is $2$ and there are $(n_1 - 1)$ such distance; the shortest path distance from a non-star node in $V_2$ is $3$ and there are $(n_2 - 1)$ such distance; the shortest distance from the star node $j$ is $1$. The sum of the shortest path distances of a non-star node $k$ in $V_1$ is $V_1\_nonStar\_distance = (2 * (n_1 - 1)) + (3 * (n_2 - 1)) + 1$,
and the closeness centrality of the non-star node $k$ will be $M_{C\_V_1\_nonStar} = \frac{2 * (n_1 - 1) + n_2}{V_1\_nonStar\_distance}$.
Each non-star node in $V_1$ exhibits the same closeness centrality value as the node $k$, and hence, there are $(n_1-1)$ such closeness centrality difference between the star node $i$ in and a non-star node in $V_1$, which is $(1-M_{C\_V_1\_nonStar})$, leading to the maximum sum of closeness measure difference in the node set $V_1$ as $Closeness_{diff\_V_1} = (n_1-1)*(1-M_{C\_V_1\_nonStar})$.
Deriving the same for the node set $V_2$ the same process, we get $V_2\_nonStar\_distance = (2 * (n_2 - 1)) + (3 * (n_1 - 1)) + 1$, $M_{C\_V_2\_nonStar} = \frac{2 * (n_2 - 1) + n_1}{V_2\_nonStar\_distance}$, and $Closeness_{diff\_V_2} = (n_2-1)*(1-M_{C\_V_2\_nonStar})$.
Finally, following the computation of Eq. 1, we have the closeness graph centralization index $C_C$ as:
\begin{math}
    \resizebox{3.5in}{!}{$
    C_C({\cal G}) = \frac{\sum_{k=1}^{n_1} [M_{C\_V_1}(p^{*}_{V_1},{\cal G})-M_{C\_V_1}(p^{ }_{V_1k},{\cal G})]}{Closeness_{diff\_V_1}}
     + \frac{\sum_{k=1}^{n_2} [M_{C\_V_2}(p^{*}_{V_2},{\cal G})-M_{C\_V_2}(p^{ }_{V_2k},{\cal G})]}{Closeness_{diff\_V_2}}$}
\end{math}
, where $M_{C\_V_1}(p^{ }_{V_1k},{\cal G}))$ can be either the in-closeness (i.e., using the incoming paths to the node $k$) or out-closeness (i.e., using the outgoing paths from the node $k$) of the node $p^{ }_{V_1k}$ in $V_1$, and so can $M_{C\_V_2}(p^{ }_{V_2k},{\cal G})$.

\end{document}